\documentstyle[twoside,fleqn,npb,epsfig]{article}

\newcommand{\AmS}{{\protect\the\textfont2
  A\kern-.1667em\lower.5ex\hbox{M}\kern-.125emS}}
\newcommand{\beq}{\begin{equation}}
\newcommand{\beqa}{\begin{eqnarray}}
\newcommand{\bce}{\begin{center}}
\newcommand{\bfig}{\begin{figure}}
\newcommand{\bit}{\begin{itemize}}
\newcommand{\ben}{\begin{enumerate}}
\newcommand{\eeq}{\end{equation}}
\newcommand{\eeqa}{\end{eqnarray}}
\newcommand{\ece}{\end{center}}
\newcommand{\efig}{\end{figure}}
\newcommand{\eit}{\end{itemize}}
\newcommand{\een}{\end{enumerate}}
\newcommand{\npb}{Nucl. Phys. {\bf B}}
\newcommand{\plb}{Phys. Lett. {\bf B}}

\newcommand{\xs}{cross-section}

\title{\rightline{\small \tt FNT/T-2000/13}
Electroweak Physics in Six-Fermion Processes at Future 
Colliders\thanks{Presented at {\sl Loops and Legs in Quantum Field Theory}, 
April 2000, Bastei, Germany. Based on work done in collaboration 
with F.~Gangemi, G.~Montagna, M.~Moretti and O.~Nicrosini.}}

\author{F. Piccinini\address{Istituto Nazionale di Fisica Nucleare, Sezione di
Pavia, \\
Dipartimento di Fisica Nucleare e Teorica, Universit\`a di Pavia, \\
Pavia, Italy}}

\begin{document}

\begin{abstract}
Recent developments in the field of complete electroweak tree-level
calculations for six-fermion final states in $e^+ e^-$ collisions are 
briefly reviewed. Particular attention is given to {\it top}-quark and 
Higgs boson physics, which are items of primary importance at the Next 
Linear Collider. The relevance of electroweak backgrounds and finite-width 
effects is discussed, showing the importance of complete calculations for 
precision studies at the colliders operating in the TeV energy range.
\end{abstract}

\maketitle

\section{Introduction}
\label{sec:intro}
Many signals of interest for tests of the Standard Model and search for new
physics at the Next Linear Collider (NLC) will be given by many-particle 
final states. 
In particular, the six-fermion signatures will be relevant to several subjects,
such as {\it top}-quark physics, intermediate-mass Higgs boson production and
the analysis of anomalous gauge couplings. 
Each of these topics can be considered as a signal by itself or as a 
background for another one. 
Assuming a realistic luminosity of $500~$fb$^{-1}$/yr, 
the statistical errors on several 
six-fermion signatures are at the per cent level, so that ``precision'' 
calculations that take into account all background, finite width and 
final-state correlation effects are needed. 
In this contribution some theoretical issues concerning six-fermion channels 
relevant for {\it top}-quark and Higgs boson physics are addressed. 
A first study of the impact of quartic anomalous gauge couplings has been 
perfomed in ref.~\cite{faban}. For the sake of brevity, 
this subject will not be reviewed in the present contribution.

The numerical results presented have been performed by means of the computer 
code {\tt SIXFAP}~\cite{sixfap}, that involves the algorithm 
{\tt ALPHA}~\cite{alpha}, for the automatic calculation of the scattering 
amplitudes, and a Monte Carlo integration procedure derived from the 
four-fermion codes {\tt HIGGSPV}~\cite{higgspv} and 
{\tt WWGENPV}~\cite{wwgenpv}, and developed to deal with six-fermion processes.
The code has been adapted to deal with a large variety of diagram topologies, 
including both charged and neutral currents, so as to keep under control all 
the relevant signals of interest as well as the complicated backgrounds 
that are involved in six-fermion processes where hundreds of diagrams 
contribute to the tree-level amplitudes. The effects of initial state 
radiation (ISR) and beamsstrahlung (BS) are also included.

\section{\bf {\it Top}-quark physics in six-quark processes}
\label{sec:top}

The production of a $t\bar t$ pair gives rise to six fermions in the
final state. The $6f$ signatures relevant to the study of the {\it top} quark
in $e^+e^-$ collisions can be summarized as follows: $b\bar bl\nu_ll'\nu_{l'}$
(leptonic, $\sim 10\%$ of the total rate), $b\bar bq\bar q'l\nu_l$ (semi
leptonic, $\sim 45\%$), $b\bar b+4q\quad$ (hadronic, $\sim 45\%$).
Semi leptonic signatures have been considered in refs.~\cite{to1,kek,4jln}.
It is then of great interest to carefully evaluate the size of the totally
hadronic, six-quark ($6q$) contributions to integrated cross-sections and
distributions as well as to determine their phenomenological 
features~\cite{6ftop,desy123f}.

The $6q$ signatures of the form $b\bar b+4q$, with $q=u,d,c,s$, are considered
and the results of complete electroweak tree-level calculations are presented. 
In particular the r\^ole of electroweak backgrounds and of ISR and BS are 
studied. The QCD contributions, which however play a significant r\^ole, are 
not considered.

In Tab.~\ref{tab:1bb} all the electroweak processes contributing to the 
signature of six quarks in the final state with one $b {\bar b}$ pair are 
listed. While the CC processes receive contribution from the $t {\bar t}$ 
production Feynman diagrams only, to the MIXED processes contribute both signal 
and background as well as their interferences diagrams. The purely NC processes 
do not contribute to the signal. 

\begin{table}[h]
\bce
\begin{tabular}[b]{p{2truecm}p{2truecm}p{2truecm}}
\hline
CC only & MIXED & NC only\\
\hline
$b\bar bu\bar d\bar cs$ & $b\bar bu\bar d\bar ud$ & $b\bar bu\bar us\bar s$,
$b\bar bc\bar cd\bar d$ \\

$b\bar b\bar udc\bar s$ & $b\bar bc\bar s\bar cs$ & $b\bar bu\bar uu\bar u$, 
$b\bar bc\bar cc\bar c$ \\

                        &                         & $b\bar bd\bar dd\bar d$, 
			$b\bar bs\bar ss\bar s$ \\

                        &                         & $b\bar bu\bar uc\bar c$ \\
                        &                         & $b\bar bd\bar ds\bar s$ \\
\hline
\end{tabular}
\caption{\small Six-quark final states with only one $b\bar b$ pair. The
notations CC (charged currents) and NC (neutral currents) refer to the currents
formed by the quark flavours different from $b$.}
\label{tab:1bb}
\ece
\end{table}

The integrated \xs\ for $e^+ e^- \to 6q $ with one $b {\bar b}$ pair 
is shown in fig.~\ref{fig:topenscan} in the energy range 
between 350 and 800 GeV for $m_h = 185$~GeV and for a realistic set of cuts 
specified in the figure. The effect of the electroweak backgrounds 
is pointed out 
by comparing the solid line, obtained with the full six-fermion calculation, 
with 
the dotted one, which corresponds to the contribution of the signal Feynman 
diagrams only. As can be seen, the contribution of the electroweak backgrounds 
reach the size of 10\% over all the center of mass (c.m.) energy spectrum. 
The results of a calculation taking into account the effect of ISR 
on the signal only are shown with the dashed-dotted line. The 
effects of ISR can be of the order of 30\% near threshold, where the Born \xs\ 
has a steep increasing, and reduce to higher energies. 

A complete six-fermion calculation can also be used 
to study the reliability of the \xs\ calculation performed in the 
narrow width approximation (NWA), 
where the {\it top}-quark is treated as a real particle, 
with the advantage of handling a simple calculation. 
As shown in ref.~\cite{6ftop} this approximation is valid at very high energy, but it 
overestimates the \xs\ of about 10\% near the threshold.

\bfig[t]
\bce
\epsfig{file=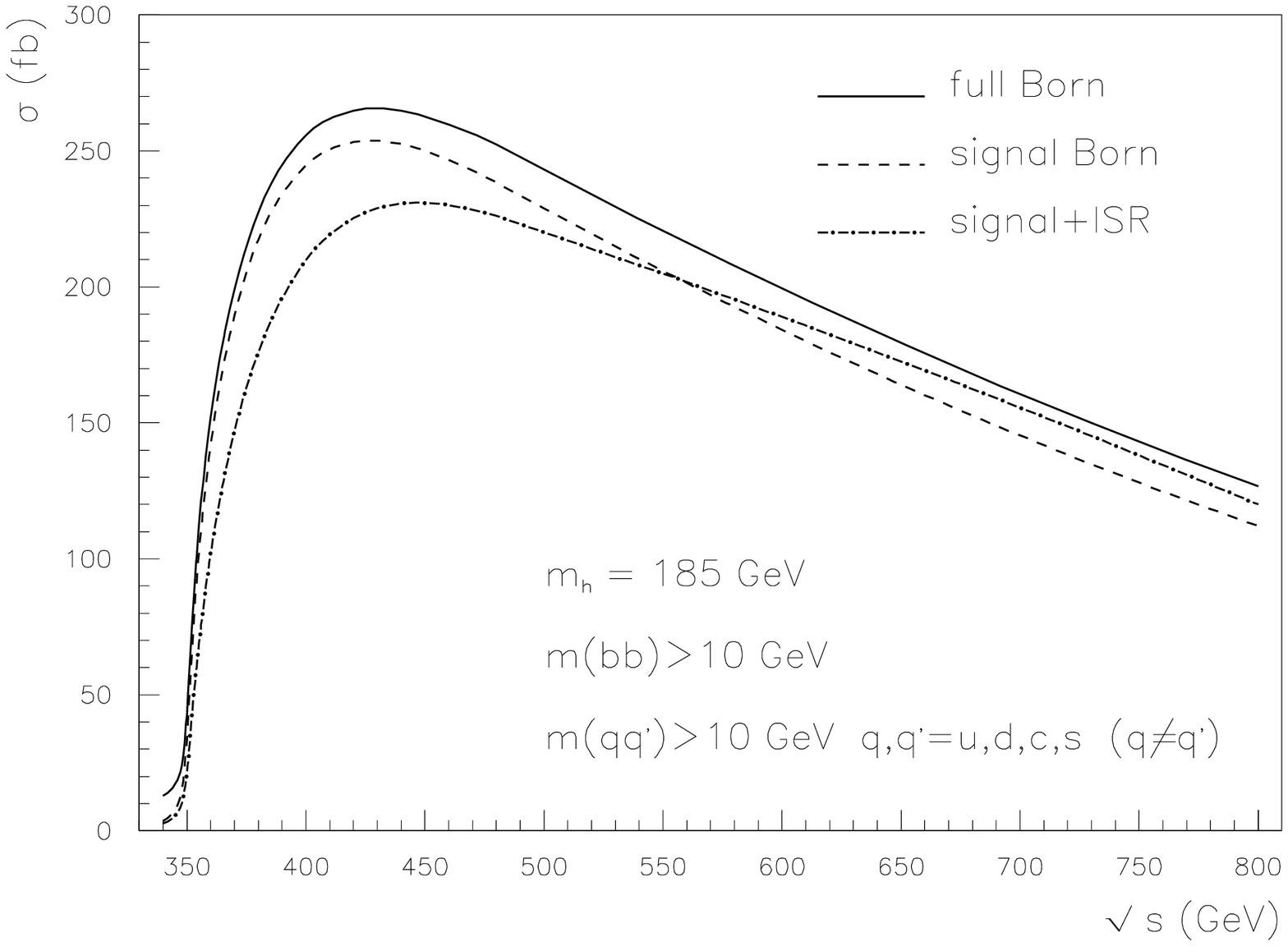,height=7cm,width=8cm}
\caption{\small Full six-quark electroweak cross section (solid line) and 
$t \bar t$ signal (dashed line) in the Born approximation, and $t \bar t$
signal with ISR (dash-dotted line), as a function of 
the c.m. energy.}
\label{fig:topenscan}
\ece
\efig

The total electroweak \xs\ has also been studied at the threshold for $t\bar t$
production as a function of the Higgs boson mass. 
Although the dominant effects in
this case come from QCD contributions, as is well known \cite{teub}, the
electroweak backgrounds turn out to give a sizeable uncertainty, of the order
of $10\%$ of the pure electroweak contribution, in the intermediate range of
Higgs boson masses (see fig.~\ref{fig:hmscan}), which is related to the fact that the
Higgs boson mass is not known. 
Furthermore, even if the Higgs boson mass value will be known 
at the run time of the NLC, its experimental error will translate into a 
theoretical uncertainty on the threshold cross-section determination, 
especially if the central value lyes in the interval 120-160~GeV.

\bfig
\bce
\epsfig{file=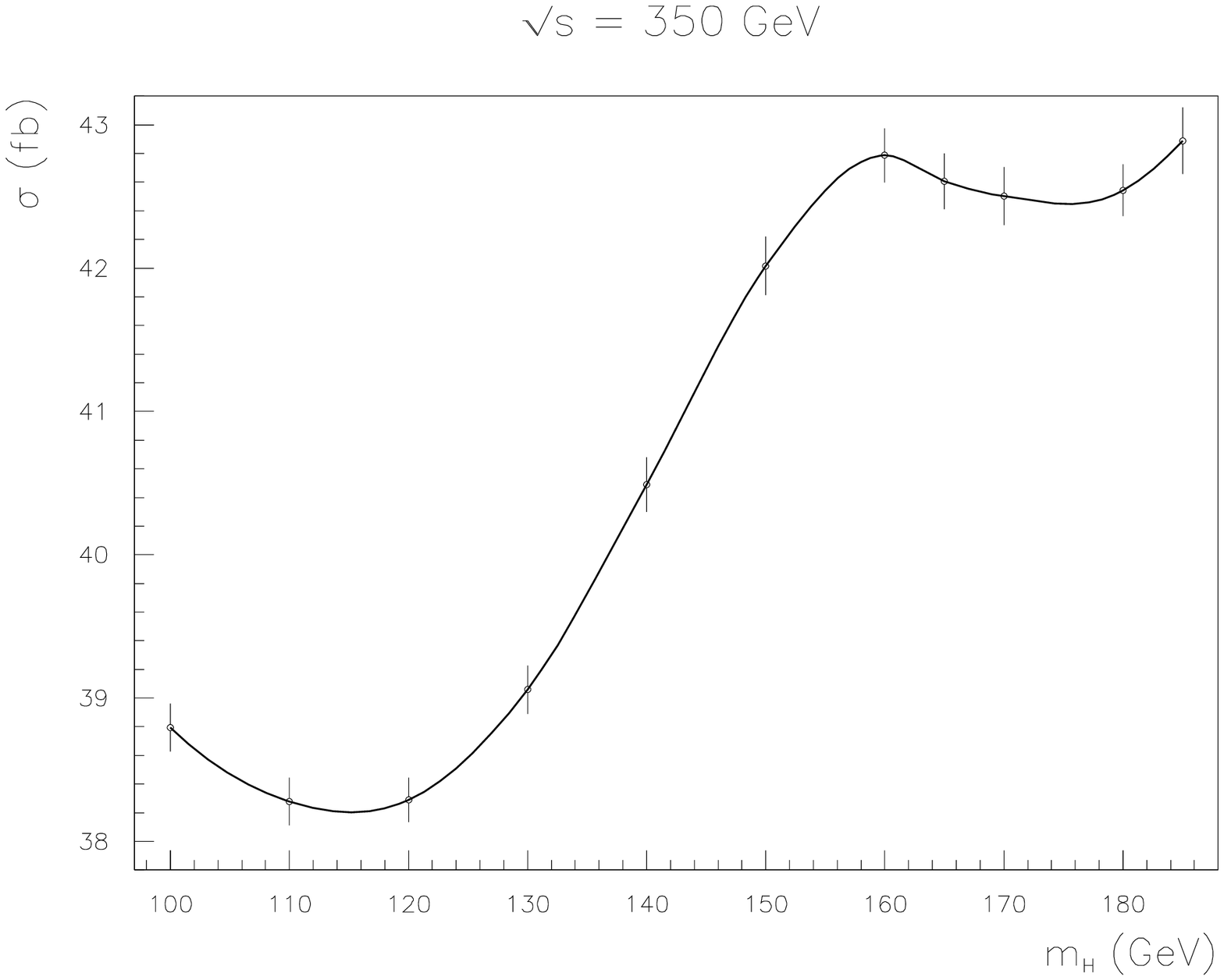,height=7cm,width=8cm}
\caption{\small Total cross-section as a function of the Higgs boson mass at 
the threshold for $t\bar t$ production.}
\label{fig:hmscan}
\ece
\efig

In order to study the possibility of isolating the {\it top}-quark
signal from the QCD backgrounds, the topology of the events can be studied 
by means of the event-shape variables. The pure QCD contributions have 
been analysed in ref.~\cite{sixj}. 
In fig.~\ref{fig:shape} the thrust distribution of the
electroweak contribution is shown at a c.m. energy of 500 GeV and with a Higgs
boson mass of 185 GeV in the Born approximation 
(dashed histogram) and with ISR and BS (solid histogram). 
The invariant masses of the $b\bar b$ pair and of all the pairs of quarks other
than $b$ are required to be greater than 10 GeV. Remarkable effects due to ISR
and BS can be seen in this plot, where the peak in the thrust distribution is
strongly reduced with respect to the Born approximation and the events are
shifted towards the lower values of $T$, which correspond to spherical events.
In view of the results of ref.~\cite{sixj}, the thrust variable is very 
effective in discriminating pure QCD backgrounds, also
in the presence of electroweak backgrounds and of ISR and BS. Going up with 
the c.m. energy, the peak in the thrust distribution approaches the 
maximum allowed value, so it becomes more difficult at energies of the order 
of 1~TeV to discriminate between electroweak contributions and 
QCD backgrounds~\cite{6ftop}.

\bfig[t]
\bce
\epsfig{file=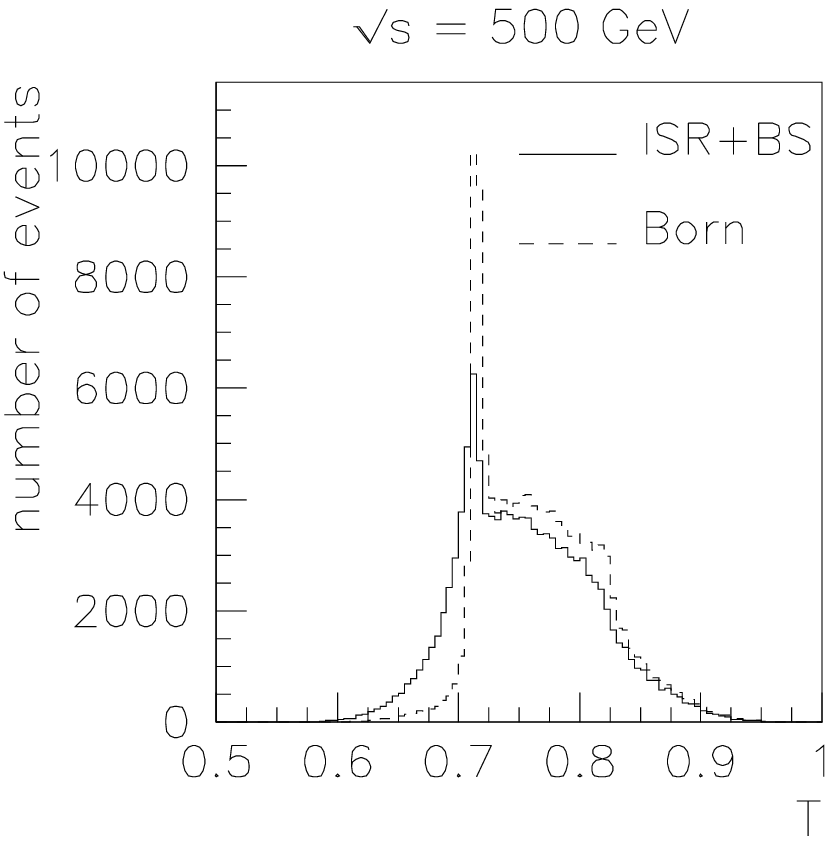,height=7.cm,width=8.cm}
\caption{\small Thrust distribution in the Born approximation (dashed histogram)
and with ISR and BS (solid histogram), at a
c.m. energy of 500 GeV and for a Higgs boson mass of 185 GeV. 
The invariant masses of the $b\bar b$ pair and of all the
pairs of quarks other than $b$ are required to be greater than 10 GeV.}
\label{fig:shape}
\ece
\efig

\section{Intermediate-mass Higgs boson}
\label{sec:Higgs}

The current lower bound on the Higgs boson mass deduced from direct search 
at LEP is 108 GeV at 95 $\%$ C.L. \cite{Quast}, 
while the upper bound given by fits to the precision data on electroweak 
observables is 188 GeV at 95 $\%$ C.L. (280 GeV at 95 $\%$ C.L. in a 
Bayesian approach)~\cite{Quast}.

In the mass range favoured by the present experimental
information, the relevant signatures at the NLC are 
four-fermion final states if the Higgs boson mass is below 130-140 GeV, and
six-fermion final states if the Higgs boson mass is greater than 140 GeV. The
processes of the first kind have been extensively studied in connection with
physics at LEP, while those of the second kind have only recently been 
addressed~\cite{to1,desy123f,sixfzpc,6fhiggs}.

In this section some aspects of complete electroweak tree-level 
calculations for the
processes $e^+e^-\to q\overline q l^+ l^- \nu\overline\nu$, with $q=u,d,c,s$,
$l=e,\mu,\tau$ and $\nu=\nu_e,\nu_\mu,\nu_\tau$ are presented. These processes
are characterized by the presence of both charged and neutral currents and of
different mechanisms of Higgs boson production involving Higgs-strahlung 
and vector boson fusion; moreover, QCD backgrounds are absent.

\bfig[h]
\bce
\epsfig{file=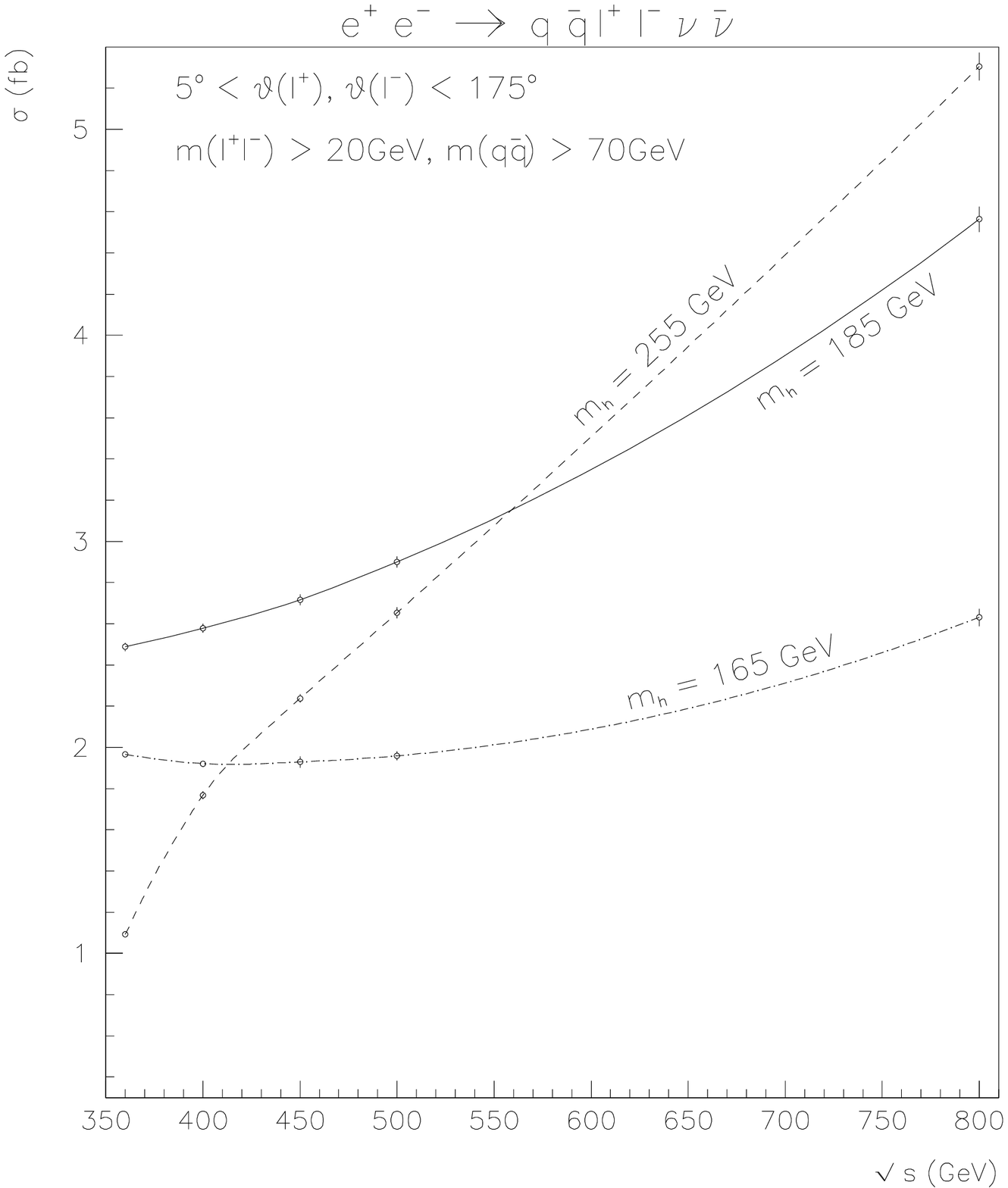,height=7cm,width=8cm}
\caption{\small Total cross section for the process 
 $e^+e^-\to q\overline ql^+l^-\nu\overline\nu$ in the Born approximation,
 as a function of $\sqrt s$ for three different values of the Higgs boson mass
$m_H$. The angles $\theta(l^+)$, $\theta(l^-)$
of the charged leptons with the beam axis are in the interval 
$5^\circ$-$175^\circ$,
the $e^+e^-$ and the $q\bar q$ invariant masses are larger than $20$ GeV.}
\label{fig:6fsscan20}
\ece
\efig

The total cross-section is shown in fig.~\ref{fig:6fsscan20} as a function of
the c.m. energy for three values of the Higgs boson mass, with
suitable kinematical cuts, to avoid the soft-pair singularities.
The increase with energy, common to all three curves in
fig.~\ref{fig:6fsscan20}, is due, at high energies, to the $t$-channel
contributions; in the case of $m_H=255$ GeV, the steep rise near
$\sqrt{s}=360$ GeV is related to the existence of a threshold
effect for the Higgs-strahlung process at an energy $\sqrt{s}\sim m_H + M_Z$.
Thanks to the sums over quark,
charged lepton and neutrino flavours, as well as the combined action of
different production mechanisms, assuming a luminosity of $500$ fb$^{-1}$/yr
and a Higgs mass of, say, $185$ GeV, more than $1000$ events can be expected
at a c.m. energy of $360$ GeV and more than $2000$ at $800$ GeV.

In the framework of a six-fermion calculation, the only meaningful procedure 
for a cross section evaluation is based on the sum of all the tree-level 
Feynman diagrams. 
On the other hand, there is a number of reasons to
consider a subset of diagrams that can be defined as the Higgs boson signal 
and to define a corresponding background. First of all, this is of great 
interest from the point of view of the search for the Higgs boson in the 
experiments. Moreover, such a definition allows one to 
make a comparison with results obtained in the NWA~\cite{hstr}--\cite{zervast},
which are the only available
estimations unless a complete $6f$ calculation is performed.
In principle, whenever a subset of diagrams is singled out, gauge invariance
may be lost and unitarity problems may arise. However, in the 
literature~\cite{6fhiggs}, an
operative definition of signal and background has been considered and its 
reliability has been studied for various Higgs boson masses and c.m. energies.

The Higgs boson signal for a given six-fermion final state is defined 
as the sum of the graphs containing a resonant Higgs boson. 
The background is defined as the sum of all
the diagrams without a Higgs boson. However, there are Feynman diagrams with 
a non-resonant Higgs boson exchanged with space-like momentum, which are
neither accounted for in the signal, nor in the background. 
Such a choice has been
dictated by the fact that these non-resonant contributions cannot correctly be
included in the signal, since they cannot find a counterpart in the NWA, and
because of gauge cancellations with background
contributions at high energies; however, as they depend on the Higgs boson 
mass, they should not be included in the background as well.
In order to give a quantitative estimate of the validity of this definition,
the total cross section (sum of all the tree-level
$6f$ Feynman diagrams) is compared in ref.~\cite{6fhiggs} with the 
incoherent sum of signal and background. 
A result of this study is that up to $500$ GeV the total cross
section and the incoherent sum of signal and background are indistinguishable
at a level of accuracy of $1\%$,
and the definition of signal may be considered meaningful; at higher
energies, the separation of signal and background starts to be less reliable,
since it requires to neglect effects that are relevant at this accuracy. In
particular, at $800$ GeV the deviation is of the order of a few per cent and it
decreases when the Higgs boson mass passes from $165$ to $185$ and to $255$ 
GeV. The results are also slightly dependent on the applied cuts on the final 
state particles.

\bfig[t]
\bce
\epsfig{file=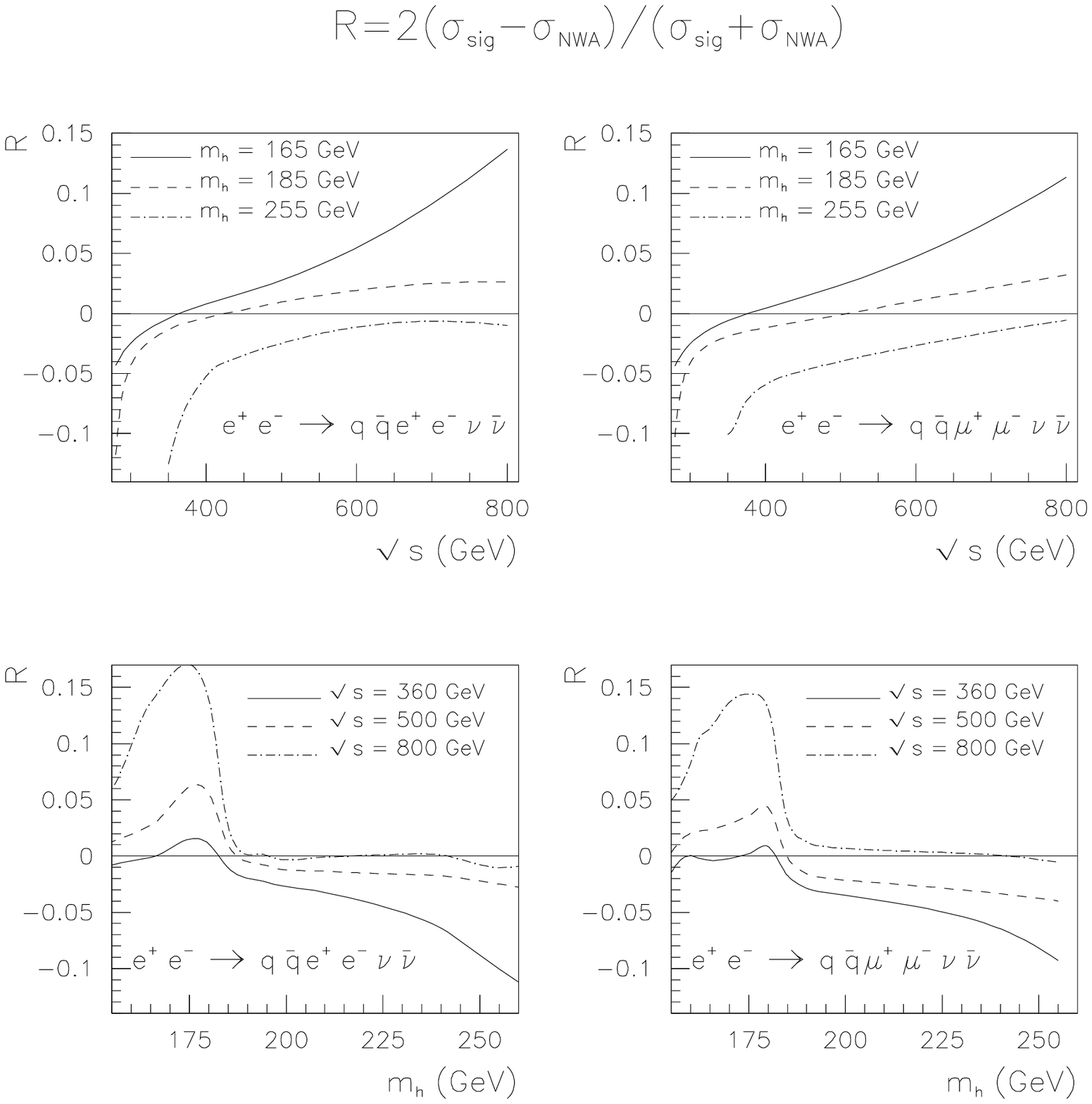,width=8cm,height=8cm}
\caption{\small Comparison between the signal cross section obtained by a
diagrammatic six-fermion calculation and the one calculated in the
NWA (see the discussion in the text),
as a function of $\sqrt s$ (upper row) and of the Higgs boson mass
(lower row).}
\label{fig:6ff-nwa}
\ece
\efig

After having tested the reliability of the Higgs boson signal definition in a
complete tree-level calculation, it is very interesting to study the accuracy 
of a theoretical prediction obtained within the NWA, 
which is much easier to perform than a full six-fermion calculation.
A comparison with the NWA is shown in 
fig.~\ref{fig:6ff-nwa} for the processes
$e^+e^-\to q\overline q e^+ e^- \nu\overline\nu$ and
$e^+e^-\to q\overline q \mu^+ \mu^- \nu\overline\nu$, where 
no kinematical cuts are applied and the results are in the Born
approximation. Here $\sigma_{sig}$
is the signal cross section, containing the contributions of the 
signal diagrams and their interferences.
The cross section in the NWA, $\sigma_{NWA}$, is obtained in the following way
(for definiteness the case with $e^+e^-$ in the final state is considered): the
known cross sections for the processes of real Higgs boson production 
$e^+e^-\to h\nu\overline\nu,he^+e^-$ ~\cite{almele,zervast} and 
$e^+e^-\to Zh$~\cite{hstr} are multiplied by the appropriate branching ratios;
then the incoherent sum of these terms is taken.
Thus the comparison between $\sigma_{sig}$ and $\sigma_{NWA}$
gives a measure of interference between the different production mechanisms
and of off-shellness effects together.
As can be seen in fig.~\ref{fig:6ff-nwa}, the relative difference $R$ is of the
order of
some per cent, depending on the Higgs boson mass and the c.m. energy; 
in some cases
it reaches values of more than $10\%$, with no substantial difference between
the two final states considered. 
\par
The size of the off-shellness effects, comparable with the ISR lowering,
indicates the importance of a full
$6f$ calculation in order to obtain sensible phenomenological
predictions.

\section{Conclusions}
\label{sec:concl}

The six-fermion final states will be among the most relevant new signatures at
future $e^+e^-$ Linear Colliders. In particular they are interesting for 
$t\bar t$ production, for Higgs boson physics in the intermediate mass range 
and for the study of anomalous gauge couplings. 
Some aspects concerning the first two items have been reviewd in this 
contribution.
The results presented have been obtained by means of 
a Monte Carlo event generator~\cite{sixfap}, developed for complete tree-level
calculations of six-fermion final states at the energies of the 
NLC, supplemented with the effects of ISR and BS. 

In this contribution the importance of complete electroweak tree-level 
calculations has been pointed out both for $t \bar t$ and for Higgs boson 
production in some specific channels. In particular the effects of 
backgrounds and finite widths have been studied and compared with the 
predictions obtained in the NWA, which are available 
in the literature. These effects turn out to be in many cases well above 
the per cent level, so they need to be taken into account in realistic 
analysis at the NLC. Moreover, the presence of a complete six-fermion final 
state generator allows to study all kinds of final-state
distributions such as invariant masses, angular correlations and event-shape
variables, that are essential both for the detection of the signals of 
interest and for the analysis of the properties of the particles under study.

\vspace{0.4truecm}
{\bf Acknowledgements}\\
The author wishes to thank the organizers, especially J.~Bl\"umlein and
T.~Riemann, for the kind invitation and the pleasant atmosphere 
during the workshop. The author is grateful to F.~Gangemi, G.~Montagna and 
O.~Nicrosini for a careful reading of the manuscript.

\end{document}